\documentstyle[12pt]{article}

\title{M\"o\ss{}bauer null redshift experiment	II}

\author{M. De Francia
	 \and
	G. Goya
	\and
	R. C. Mercader
	\and
	H. Vucetich\\
	\it Departamento de F\'\i{}sica; Facultad de Ciencias Exactas;\\
	\it	Universidad Nacional de La Plata. CC-67 1900-La Plata\\
	\it	ARGENTINA
	}

\begin{document}

\maketitle

\begin{abstract}
	Accurate limits for the violation of the Principle of
	Equivalence are found by comparing the redshifts of
	$^{57}$Fe in different chemical environments.
\end{abstract}

	The Principle of Equivalence is the basis of gravitational theory.
It states that a freely falling frame is locally equivalent to an inertial
frame \cite{Einstein1911}. Its main consequences, that can be used to verify
it, are \cite{Will81,Will84}:
\begin{description}
  \item[Universality of Free Fall:]
	The lines of universe of test bodies are independent of their
internal structure and composition.
  \item[Local Lorentz Invariance:]
	Local non gravitational phenomena are independent of the reference
frame velocity.
  \item[Local Position Invariance:]
	Local non gravitational phenomena are independent of the origin of
the reference
frame.
\end{description}

	The simplest test theory of the above statements refers to a single
(composite) test body in a background gravitational field. It is defined by
the Lagrangian \cite{Haughan79}:
\begin{equation}
L = -m_R c^2 + \frac{1}{2}m_I {\bf v}^2 - m_P U({\bf x})
		+ O({v^4\over{c^2}}) 			\label{Lag}
\end{equation}
In this equation ${\bf x}$ and ${\bf v}$ are the center of mass coordinate
and velocity of the test body, $U({\bf x})$ its gravitational potential and
$m_R$, $m_I$ and $m_P$ are parameters called the rest, inertial and passive
gravitational masses of the test body, respectively.

	The Principle of Equivalence imposes severe restrictions on the
adjustable parameters of equation (\ref{Lag}). Universality of Free Fall
implies:
\begin{equation}
m_I = m_P,					\label{mi=mp}
\end{equation}
while Local Position Invariance requires:
\begin{equation}
m_R = m_P					\label{mr=mp}
\end{equation}
and  finally, Local Lorentz Invariance demands:
\begin{equation}
m_R = m_I					\label{mr=mi}
\end{equation}

	These three equations can be used to verify (more accurately, to
falsify) the Principle of Equivalence.
	In this letter we obtain bounds on the violation of Local Lorentz
Invariance and Local Position Invariance from equations (\ref{mr=mp}) and
(\ref{mr=mi}); testing for the existence of anomalous redshifts in a
M\"o\ss{}bauer emitter-absorber experiment \cite{Red1}.

	Consider an atomic system suffering an electromagnetic transition
between two levels $E_0$ and $ E_1$, with an energy difference $\Delta E =
E_1 - E_0$. A change of position in a gravitational field produces a
frequency shift, the gravitational redshift \cite{Haughan79,Red1}:
\begin{equation}
z_g = {{\omega-\omega'}\over{\omega}} = (1 + \xi_P){\Delta U\over c^2}
						\label{gRed}
\end{equation}
with
\begin{equation}
\xi_P = {{[\delta m_P(E_2)-\delta m_P(E_1)] c^2}\over{\Delta E}}
							\label{xi-g}
\end{equation}
 where $\delta m_P$ are the differences between the gravitational and rest
masses of the upper and lower energy levels. The first term in (\ref{gRed})
is the universal gravitational redshift predicted by the Principle of
Equivalence, while the second term is an anomalous, composition dependent
redshift \cite{Haughan79,Red1}.

	In a similar way, if the system is boosted with velocity ${\bf v}$,
its frequency will be shifted by the amount (the transverse D\"oppler shift)
\cite{Haughan79}:
\begin{equation}
z_i = -\frac{1}{2} \frac{v^2}{c^2} + \xi_I \frac{ (-\frac{1}{2} v^2
	+ {\bf W.v})}{c^2}			\label{iRed}
\end{equation}
where {\bf W} is the initial velocity of the atomic system with respect
to a privileged (``absolute'') reference frame, and
\begin{equation}
\xi_I = {{[\delta m_I(E_2)-\delta m_I(E_1)] c^2}\over{\Delta E}}
							\label{xi-i}
\end{equation}
where  $\delta m_I$ are the differences between inertial and rest masses.

	In a null redshift experiment, the redshifts of two different atomic
systems, $A$ and $B$, are compared. Only the ``anomalous'' terms in
equations (\ref{gRed}) and (\ref{iRed}) would contribute to the differential
redshift:
\begin{eqnarray}
\Delta z & = & [\xi_P(A) - \xi_P(B)]\frac{\Delta U}{c^2} +	\nonumber\\
	&   & [\xi_I(A) - \xi_I(B)]\frac{ (-\frac{1}{2} v^2
	+ {\bf W.v})}{c^2}			\label{diffRed}
\end{eqnarray}
and so, a null redshift experiment is a test for the universality of the
redshifts.

	The M\"o\ss{}bauer null redshift experiments test the difference of
redshift of the same nuclear species (in our case, $^{57}$Fe) in different
chemical environments. At first sight one expects a null relative shift,
since we are comparing the rates of two identical clocks. This is not so,
however: in the resonant emission-absorption phenomenon, the transition
occurs between two collective states of the crystal, of extremely big masses
\cite{Red1}. The mass differences $\delta m_P$ and $\delta m_I$ are fuctions
of the chemical composition of the system having, in general, nonzero values.

	Consider now a resonant emitter-absorber system moving with the
Earth along orbit. Because of its eccentricity, the Sun gravitational
potential at the laboratory will have a sinusoidal variation with the period
of one year. The same seasonal periodicity but with a different phase is
expected for the redshift due to the ``\ae{}ther wind'' term in equation
(\ref{iRed}). This characteristic signature of the breakdown of the
Principle of Equivalence, was used to seek for an anomalous redshift in the
previous experiment
\cite{Red1}. The differential anomalous redshift would appear as a yearly
fluctuation of the isomer shift between the two samples. Since the
zero-velocity channel of the spectrometer is obtained from the calibrated
difference between the isomer shifts of a standard emitter-absorber pair, a
variable isomer shift would induce a sinusoidal variation of the
zero-velocity channel of the spectrometer:  %
\begin{equation}
v(t) = v_0 + \Delta \xi_P \frac{\Delta U}{c^2} + \Delta \xi_I \frac{
	(-\frac{1}{2} v^2 + {\bf W.v}) }{c^2}
					\label{IsomerShift}
\end{equation}

\begin{figure}[p]
\caption{\label{Fig} Zero-velocity channel {\em vs} time}

% GNUPLOT: LaTeX picture
\setlength{\unitlength}{0.240900pt}
\ifx\plotpoint\undefined\newsavebox{\plotpoint}\fi
\sbox{\plotpoint}{\rule[-0.175pt]{0.350pt}{0.350pt}}%
\begin{picture}(1500,900)(0,0)
\tenrm
\sbox{\plotpoint}{\rule[-0.175pt]{0.350pt}{0.350pt}}%
\put(264,697){\rule[-0.175pt]{282.335pt}{0.350pt}}
\put(264,158){\rule[-0.175pt]{4.818pt}{0.350pt}}
\put(242,158){\makebox(0,0)[r]{-0.012}}
\put(1416,158){\rule[-0.175pt]{4.818pt}{0.350pt}}
\put(264,248){\rule[-0.175pt]{4.818pt}{0.350pt}}
\put(242,248){\makebox(0,0)[r]{-0.01}}
\put(1416,248){\rule[-0.175pt]{4.818pt}{0.350pt}}
\put(264,338){\rule[-0.175pt]{4.818pt}{0.350pt}}
\put(242,338){\makebox(0,0)[r]{-0.008}}
\put(1416,338){\rule[-0.175pt]{4.818pt}{0.350pt}}
\put(264,428){\rule[-0.175pt]{4.818pt}{0.350pt}}
\put(242,428){\makebox(0,0)[r]{-0.006}}
\put(1416,428){\rule[-0.175pt]{4.818pt}{0.350pt}}
\put(264,517){\rule[-0.175pt]{4.818pt}{0.350pt}}
\put(242,517){\makebox(0,0)[r]{-0.004}}
\put(1416,517){\rule[-0.175pt]{4.818pt}{0.350pt}}
\put(264,607){\rule[-0.175pt]{4.818pt}{0.350pt}}
\put(242,607){\makebox(0,0)[r]{-0.002}}
\put(1416,607){\rule[-0.175pt]{4.818pt}{0.350pt}}
\put(264,697){\rule[-0.175pt]{4.818pt}{0.350pt}}
\put(242,697){\makebox(0,0)[r]{0}}
\put(1416,697){\rule[-0.175pt]{4.818pt}{0.350pt}}
\put(264,787){\rule[-0.175pt]{4.818pt}{0.350pt}}
\put(242,787){\makebox(0,0)[r]{0.002}}
\put(1416,787){\rule[-0.175pt]{4.818pt}{0.350pt}}
\put(264,158){\rule[-0.175pt]{0.350pt}{4.818pt}}
\put(264,113){\makebox(0,0){1984}}
\put(264,767){\rule[-0.175pt]{0.350pt}{4.818pt}}
\put(394,158){\rule[-0.175pt]{0.350pt}{4.818pt}}
\put(394,113){\makebox(0,0){1985}}
\put(394,767){\rule[-0.175pt]{0.350pt}{4.818pt}}
\put(524,158){\rule[-0.175pt]{0.350pt}{4.818pt}}
\put(524,113){\makebox(0,0){1986}}
\put(524,767){\rule[-0.175pt]{0.350pt}{4.818pt}}
\put(655,158){\rule[-0.175pt]{0.350pt}{4.818pt}}
\put(655,113){\makebox(0,0){1987}}
\put(655,767){\rule[-0.175pt]{0.350pt}{4.818pt}}
\put(785,158){\rule[-0.175pt]{0.350pt}{4.818pt}}
\put(785,113){\makebox(0,0){1988}}
\put(785,767){\rule[-0.175pt]{0.350pt}{4.818pt}}
\put(915,158){\rule[-0.175pt]{0.350pt}{4.818pt}}
\put(915,113){\makebox(0,0){1989}}
\put(915,767){\rule[-0.175pt]{0.350pt}{4.818pt}}
\put(1045,158){\rule[-0.175pt]{0.350pt}{4.818pt}}
\put(1045,113){\makebox(0,0){1990}}
\put(1045,767){\rule[-0.175pt]{0.350pt}{4.818pt}}
\put(1176,158){\rule[-0.175pt]{0.350pt}{4.818pt}}
\put(1176,113){\makebox(0,0){1991}}
\put(1176,767){\rule[-0.175pt]{0.350pt}{4.818pt}}
\put(1306,158){\rule[-0.175pt]{0.350pt}{4.818pt}}
\put(1306,113){\makebox(0,0){1992}}
\put(1306,767){\rule[-0.175pt]{0.350pt}{4.818pt}}
\put(1436,158){\rule[-0.175pt]{0.350pt}{4.818pt}}
\put(1436,113){\makebox(0,0){1993}}
\put(1436,767){\rule[-0.175pt]{0.350pt}{4.818pt}}
\put(264,158){\rule[-0.175pt]{282.335pt}{0.350pt}}
\put(1436,158){\rule[-0.175pt]{0.350pt}{151.526pt}}
\put(264,787){\rule[-0.175pt]{282.335pt}{0.350pt}}
\put(1,472){\makebox(0,0)[l]{\shortstack{$v_i$\\{}[mm/s]}}}
\put(850,68){\makebox(0,0){$t$ [yr]}}
\put(264,158){\rule[-0.175pt]{0.350pt}{151.526pt}}
\put(1350,722){\raisebox{-1.2pt}{\makebox(0,0){$\Diamond$}}}
\put(379,590){\raisebox{-1.2pt}{\makebox(0,0){$\Diamond$}}}
\put(401,535){\raisebox{-1.2pt}{\makebox(0,0){$\Diamond$}}}
\put(331,527){\raisebox{-1.2pt}{\makebox(0,0){$\Diamond$}}}
\put(396,521){\raisebox{-1.2pt}{\makebox(0,0){$\Diamond$}}}
\put(444,588){\raisebox{-1.2pt}{\makebox(0,0){$\Diamond$}}}
\put(359,571){\raisebox{-1.2pt}{\makebox(0,0){$\Diamond$}}}
\put(407,587){\raisebox{-1.2pt}{\makebox(0,0){$\Diamond$}}}
\put(392,575){\raisebox{-1.2pt}{\makebox(0,0){$\Diamond$}}}
\put(440,392){\raisebox{-1.2pt}{\makebox(0,0){$\Diamond$}}}
\put(457,573){\raisebox{-1.2pt}{\makebox(0,0){$\Diamond$}}}
\put(420,485){\raisebox{-1.2pt}{\makebox(0,0){$\Diamond$}}}
\put(429,551){\raisebox{-1.2pt}{\makebox(0,0){$\Diamond$}}}
\put(477,456){\raisebox{-1.2pt}{\makebox(0,0){$\Diamond$}}}
\put(420,600){\raisebox{-1.2pt}{\makebox(0,0){$\Diamond$}}}
\put(414,486){\raisebox{-1.2pt}{\makebox(0,0){$\Diamond$}}}
\put(483,529){\raisebox{-1.2pt}{\makebox(0,0){$\Diamond$}}}
\put(488,658){\raisebox{-1.2pt}{\makebox(0,0){$\Diamond$}}}
\put(496,526){\raisebox{-1.2pt}{\makebox(0,0){$\Diamond$}}}
\put(446,407){\raisebox{-1.2pt}{\makebox(0,0){$\Diamond$}}}
\put(494,573){\raisebox{-1.2pt}{\makebox(0,0){$\Diamond$}}}
\put(442,642){\raisebox{-1.2pt}{\makebox(0,0){$\Diamond$}}}
\put(472,612){\raisebox{-1.2pt}{\makebox(0,0){$\Diamond$}}}
\put(511,511){\raisebox{-1.2pt}{\makebox(0,0){$\Diamond$}}}
\put(542,517){\raisebox{-1.2pt}{\makebox(0,0){$\Diamond$}}}
\put(555,642){\raisebox{-1.2pt}{\makebox(0,0){$\Diamond$}}}
\put(462,725){\raisebox{-1.2pt}{\makebox(0,0){$\Diamond$}}}
\put(564,641){\raisebox{-1.2pt}{\makebox(0,0){$\Diamond$}}}
\put(568,353){\raisebox{-1.2pt}{\makebox(0,0){$\Diamond$}}}
\put(566,516){\raisebox{-1.2pt}{\makebox(0,0){$\Diamond$}}}
\put(574,629){\raisebox{-1.2pt}{\makebox(0,0){$\Diamond$}}}
\put(596,479){\raisebox{-1.2pt}{\makebox(0,0){$\Diamond$}}}
\put(583,539){\raisebox{-1.2pt}{\makebox(0,0){$\Diamond$}}}
\put(535,625){\raisebox{-1.2pt}{\makebox(0,0){$\Diamond$}}}
\put(544,491){\raisebox{-1.2pt}{\makebox(0,0){$\Diamond$}}}
\put(520,522){\raisebox{-1.2pt}{\makebox(0,0){$\Diamond$}}}
\put(624,528){\raisebox{-1.2pt}{\makebox(0,0){$\Diamond$}}}
\put(600,509){\raisebox{-1.2pt}{\makebox(0,0){$\Diamond$}}}
\put(577,560){\raisebox{-1.2pt}{\makebox(0,0){$\Diamond$}}}
\put(581,479){\raisebox{-1.2pt}{\makebox(0,0){$\Diamond$}}}
\put(570,557){\raisebox{-1.2pt}{\makebox(0,0){$\Diamond$}}}
\put(711,479){\raisebox{-1.2pt}{\makebox(0,0){$\Diamond$}}}
\put(674,425){\raisebox{-1.2pt}{\makebox(0,0){$\Diamond$}}}
\put(735,562){\raisebox{-1.2pt}{\makebox(0,0){$\Diamond$}}}
\put(698,446){\raisebox{-1.2pt}{\makebox(0,0){$\Diamond$}}}
\put(724,399){\raisebox{-1.2pt}{\makebox(0,0){$\Diamond$}}}
\put(728,606){\raisebox{-1.2pt}{\makebox(0,0){$\Diamond$}}}
\put(648,533){\raisebox{-1.2pt}{\makebox(0,0){$\Diamond$}}}
\put(666,637){\raisebox{-1.2pt}{\makebox(0,0){$\Diamond$}}}
\put(696,453){\raisebox{-1.2pt}{\makebox(0,0){$\Diamond$}}}
\put(700,591){\raisebox{-1.2pt}{\makebox(0,0){$\Diamond$}}}
\put(735,486){\raisebox{-1.2pt}{\makebox(0,0){$\Diamond$}}}
\put(752,437){\raisebox{-1.2pt}{\makebox(0,0){$\Diamond$}}}
\put(663,509){\raisebox{-1.2pt}{\makebox(0,0){$\Diamond$}}}
\put(768,461){\raisebox{-1.2pt}{\makebox(0,0){$\Diamond$}}}
\put(689,495){\raisebox{-1.2pt}{\makebox(0,0){$\Diamond$}}}
\put(785,527){\raisebox{-1.2pt}{\makebox(0,0){$\Diamond$}}}
\put(798,609){\raisebox{-1.2pt}{\makebox(0,0){$\Diamond$}}}
\put(757,474){\raisebox{-1.2pt}{\makebox(0,0){$\Diamond$}}}
\put(770,333){\raisebox{-1.2pt}{\makebox(0,0){$\Diamond$}}}
\put(702,469){\raisebox{-1.2pt}{\makebox(0,0){$\Diamond$}}}
\put(789,429){\raisebox{-1.2pt}{\makebox(0,0){$\Diamond$}}}
\put(726,481){\raisebox{-1.2pt}{\makebox(0,0){$\Diamond$}}}
\put(731,501){\raisebox{-1.2pt}{\makebox(0,0){$\Diamond$}}}
\put(791,557){\raisebox{-1.2pt}{\makebox(0,0){$\Diamond$}}}
\put(822,626){\raisebox{-1.2pt}{\makebox(0,0){$\Diamond$}}}
\put(733,464){\raisebox{-1.2pt}{\makebox(0,0){$\Diamond$}}}
\put(772,620){\raisebox{-1.2pt}{\makebox(0,0){$\Diamond$}}}
\put(757,687){\raisebox{-1.2pt}{\makebox(0,0){$\Diamond$}}}
\put(822,594){\raisebox{-1.2pt}{\makebox(0,0){$\Diamond$}}}
\put(852,742){\raisebox{-1.2pt}{\makebox(0,0){$\Diamond$}}}
\put(759,742){\raisebox{-1.2pt}{\makebox(0,0){$\Diamond$}}}
\put(785,451){\raisebox{-1.2pt}{\makebox(0,0){$\Diamond$}}}
\put(876,450){\raisebox{-1.2pt}{\makebox(0,0){$\Diamond$}}}
\put(774,316){\raisebox{-1.2pt}{\makebox(0,0){$\Diamond$}}}
\put(804,517){\raisebox{-1.2pt}{\makebox(0,0){$\Diamond$}}}
\put(874,540){\raisebox{-1.2pt}{\makebox(0,0){$\Diamond$}}}
\put(852,570){\raisebox{-1.2pt}{\makebox(0,0){$\Diamond$}}}
\put(828,418){\raisebox{-1.2pt}{\makebox(0,0){$\Diamond$}}}
\put(876,501){\raisebox{-1.2pt}{\makebox(0,0){$\Diamond$}}}
\put(841,554){\raisebox{-1.2pt}{\makebox(0,0){$\Diamond$}}}
\put(846,532){\raisebox{-1.2pt}{\makebox(0,0){$\Diamond$}}}
\put(880,422){\raisebox{-1.2pt}{\makebox(0,0){$\Diamond$}}}
\put(885,586){\raisebox{-1.2pt}{\makebox(0,0){$\Diamond$}}}
\put(863,487){\raisebox{-1.2pt}{\makebox(0,0){$\Diamond$}}}
\put(885,655){\raisebox{-1.2pt}{\makebox(0,0){$\Diamond$}}}
\put(896,531){\raisebox{-1.2pt}{\makebox(0,0){$\Diamond$}}}
\put(887,665){\raisebox{-1.2pt}{\makebox(0,0){$\Diamond$}}}
\put(945,584){\raisebox{-1.2pt}{\makebox(0,0){$\Diamond$}}}
\put(948,424){\raisebox{-1.2pt}{\makebox(0,0){$\Diamond$}}}
\put(972,489){\raisebox{-1.2pt}{\makebox(0,0){$\Diamond$}}}
\put(950,407){\raisebox{-1.2pt}{\makebox(0,0){$\Diamond$}}}
\put(1034,534){\raisebox{-1.2pt}{\makebox(0,0){$\Diamond$}}}
\put(980,590){\raisebox{-1.2pt}{\makebox(0,0){$\Diamond$}}}
\put(1008,239){\raisebox{-1.2pt}{\makebox(0,0){$\Diamond$}}}
\put(1050,577){\raisebox{-1.2pt}{\makebox(0,0){$\Diamond$}}}
\put(1063,345){\raisebox{-1.2pt}{\makebox(0,0){$\Diamond$}}}
\put(1039,671){\raisebox{-1.2pt}{\makebox(0,0){$\Diamond$}}}
\put(1100,614){\raisebox{-1.2pt}{\makebox(0,0){$\Diamond$}}}
\put(1093,343){\raisebox{-1.2pt}{\makebox(0,0){$\Diamond$}}}
\put(1082,230){\raisebox{-1.2pt}{\makebox(0,0){$\Diamond$}}}
\put(1093,395){\raisebox{-1.2pt}{\makebox(0,0){$\Diamond$}}}
\put(1154,468){\raisebox{-1.2pt}{\makebox(0,0){$\Diamond$}}}
\put(1184,557){\raisebox{-1.2pt}{\makebox(0,0){$\Diamond$}}}
\put(1117,596){\raisebox{-1.2pt}{\makebox(0,0){$\Diamond$}}}
\put(1204,422){\raisebox{-1.2pt}{\makebox(0,0){$\Diamond$}}}
\put(1208,597){\raisebox{-1.2pt}{\makebox(0,0){$\Diamond$}}}
\put(1126,432){\raisebox{-1.2pt}{\makebox(0,0){$\Diamond$}}}
\put(1134,476){\raisebox{-1.2pt}{\makebox(0,0){$\Diamond$}}}
\put(1221,656){\raisebox{-1.2pt}{\makebox(0,0){$\Diamond$}}}
\put(1228,546){\raisebox{-1.2pt}{\makebox(0,0){$\Diamond$}}}
\put(1186,438){\raisebox{-1.2pt}{\makebox(0,0){$\Diamond$}}}
\put(1206,421){\raisebox{-1.2pt}{\makebox(0,0){$\Diamond$}}}
\put(1199,660){\raisebox{-1.2pt}{\makebox(0,0){$\Diamond$}}}
\put(1267,483){\raisebox{-1.2pt}{\makebox(0,0){$\Diamond$}}}
\put(1284,759){\raisebox{-1.2pt}{\makebox(0,0){$\Diamond$}}}
\put(1221,417){\raisebox{-1.2pt}{\makebox(0,0){$\Diamond$}}}
\put(1269,456){\raisebox{-1.2pt}{\makebox(0,0){$\Diamond$}}}
\put(1245,578){\raisebox{-1.2pt}{\makebox(0,0){$\Diamond$}}}
\put(1223,557){\raisebox{-1.2pt}{\makebox(0,0){$\Diamond$}}}
\put(1234,591){\raisebox{-1.2pt}{\makebox(0,0){$\Diamond$}}}
\put(1286,419){\raisebox{-1.2pt}{\makebox(0,0){$\Diamond$}}}
\put(1288,494){\raisebox{-1.2pt}{\makebox(0,0){$\Diamond$}}}
\put(1356,679){\raisebox{-1.2pt}{\makebox(0,0){$\Diamond$}}}
\put(1369,405){\raisebox{-1.2pt}{\makebox(0,0){$\Diamond$}}}
\put(1377,563){\raisebox{-1.2pt}{\makebox(0,0){$\Diamond$}}}
\put(1354,586){\raisebox{-1.2pt}{\makebox(0,0){$\Diamond$}}}
\put(1380,565){\raisebox{-1.2pt}{\makebox(0,0){$\Diamond$}}}
\put(1388,671){\raisebox{-1.2pt}{\makebox(0,0){$\Diamond$}}}
\put(1282,520){\raisebox{-1.2pt}{\makebox(0,0){$\Diamond$}}}
\end{picture}

\end{figure}
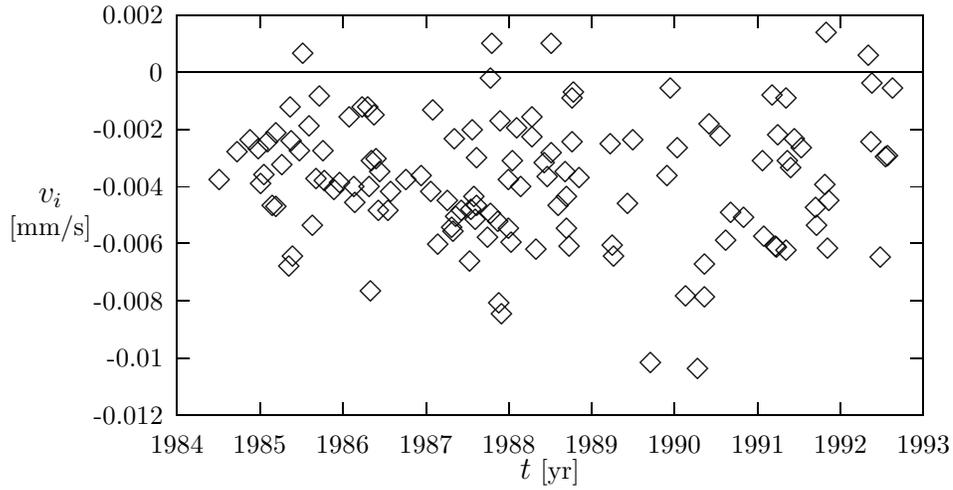

 In this experiment we have analyzed data taken
 over a period of nine years, to search for any such yearly periodic signal.
The
M\"o\ss{}bauer spectra were taken with
a natural $\alpha$-Fe foil 6 $\mu$m thick  absorber in a conventional
512 channel, constant acceleration spectrometer, with transmission geometry
\cite{Shenay}. The sources used
over this time were nominally 25 mCi $^{57}$Co in Rh matrices.  The
laboratory temperature stability was of $\pm2\;^\circ$C. The magnetically
split spectra were fitted with a nonlinear least-squares program to sextets
of lorentzians with equal linewidths for each component.  The zero-velocity
channel position was determined with respect to the centroid of the
spectrum. The baseline was simulated with a second order polynomial to
account for cosine smearing. The velocity calibration was obtained by means
of the measured $\alpha$-Fe hyperfine field  splitting.  Figure \ref{Fig}
shows the zero-velocity channel position $v(t)$ (in mm/s) as a function of
time.  Although the data were taken for calibration purposes, the absolute
values obtained are not relevant, since we are interested only in their
internal consistency.

	The time dependence of the zero-velocity channel data was adjusted
to equation (\ref{IsomerShift}) with a linear least-squares program, using
equal weights.  For {\bf W}, the velocity of the Solar System derived from
the Cosmic Microwave Background dipole was taken. All terms up to linear in
the eccentricity of the Earth orbit were kept. The resulting parameter
estimates are:  %
\begin{eqnarray}
	v_0 & = & (-3.9\pm0.2)\times10^{-3}\;{\rm mm/s}	\label{x0e}\\
	\Delta\xi_P & = & (1.5\pm1.9)\times10^{-5}	\label{xi-Pe}\\
	\Delta\xi_I & = & (-0.7\pm2.7)\times10^{-8}	\label{xi-Ie}
\end{eqnarray}

	A Kolmogorov-Smirnov test applied to the relative residuals show
they are gaussian with probability $P=0.86$. Different weighting schemes,
based on statistical counting errors, were also tried, but the
Kolmogorov-Smirnov test showed that the corresponding residuals are not
gaussian. From this we conclude that the error is dominated by random
fluctuations in the zero-velocity channel, possibly of mechanical or thermal
origins.

	Our results (\ref{x0e}) to (\ref{xi-Ie}) can be used to set upper
bounds for any violation of the Principle of Equivalence.  Table
\ref{TabI} shows the 95\% confidence limit upper bounds to $\Delta\xi_P$ and
$\Delta\xi_I$, derived from (\ref{xi-Pe}) and (\ref{xi-Ie}), together with
the results of other previous null redshift results. It can be seen that the
present work  reproduces the bounds on $\Delta\xi_P$ previously
obtained. To our knowledge, the upper bound on $\Delta\xi_I$
is the first ever obtained, since the Turner-Hill experiment
\cite{Turner-Hill} is an absolute experiment \cite{Red1}, that puts a bound
on $\xi_I$.

	We can combine our results (\ref{xi-Pe}) and (\ref{xi-Ie}) to test
the Universality of Free Fall. The difference of acceleration between the
upper and lower level in different chemical environments will certainly be
bound by the quantity:
\begin{equation}
\eta=\frac{\Delta E}{M}|(\Delta\xi_P-\Delta\xi_I)|
\end{equation}
where $M$ is the mass of a $^{57}$Fe atom, much smaller than the mass of
the  collective states. From the upper bounds in table \ref{TabI} we obtain:
\begin{equation}
|\eta| < 10^{-11}			\label{TestUFF}
\end{equation}
which is comparable to the results of modern versions of the classical
E\"otv\"os experiment
\cite{Will81,Will84,Heckel89,Heckel92}.

	We can also use the former results to find bounds for the scalar (or
pseudoscalar) fields in theories having a pseudoscalar coupling with
electromagnetism \cite{Ni,CGJ,HS}. In these theories, the coupling of the
new field $\phi$, in the form of a Chern-Simmons term:
\begin{equation}
{\cal L}_I = \kappa\phi {\bf E.B}
\end{equation}
 is assumed.  The
	Gauss law in such theories is modified in the form:
\begin{equation}
\nabla.{\bf E} = 4\pi(\rho + \kappa\nabla\phi.{\bf B})
\end{equation}
which implies the existence of a charge (pseudo)density originated by the
magnetic field:
\begin{equation}
\delta\rho_B =  \kappa\nabla\phi.{\bf B}
\end{equation}
in the $\alpha$-Fe absorber.  This charge density will
contribute with $\delta v = \alpha \delta\rho/e$ (with $e$, the electron
charge) to the isomer shift \cite{DK78}. Since the sample is not polarized,
different domains have randomly oriented magnetization and an upper bound
for the root mean square value of $\kappa\nabla\phi$ can be found from:  %
\begin{equation}
|\kappa\nabla\phi| < 3\frac{e}{B}\frac{\sigma(v_0)}{\alpha}
\end{equation}

	Using the typical values for iron ($\alpha\sim0.3$ a.u.-mm/s;
$B=330$ kgauss) \cite{Ingalls78} and the estimate of
$\sigma(v_0)$ from (\ref{x0e}) we find
\begin{equation}
|\kappa\nabla\phi| < 2\times10^5\; {\rm cm}^{-1}	\label{BoundPhi}
\end{equation}
an enormous value, compared with the estimates coming from cosmology
\cite{CGJ,HS}:
\begin{equation}
|\kappa\nabla\phi| < 10^{-28}\; {\rm cm}^{-1}		\label{BoundCos}
\end{equation}

	Our result (\ref{BoundPhi}), however, is a strictly local bound,
based only in laboratory, short time data. The factor $\sim 10^{33}$ between
equations (\ref{BoundPhi}) and (\ref{BoundCos}) comes mainly from the factor
$\sim 10^{41}$ between the
distances used to estimate the gradient: $\sim1$ fm and $\sim 10^8$ pc
respectively.

	In conclusion: the present experiment is an accurate test of the
Principle of Equivalence.  Equations (\ref{xi-Pe}), (\ref{xi-Ie}) and
(\ref{TestUFF}) provide rigorous upper bounds on the adjustable parameters
of Lagrangian (\ref{Lag}), and in turn they provide accurate tests for the
validity of the three main consequences of the Principle of Equivalence.
Besides, being carried out in systems in a strong magnetic field, it
provides a local, short time bound on the gradient of any (pseudo)scalar
field coupling to electromagnetism though a Chern-Simmons-like term.

	The authors whish to acknowledge support from CONICET  for
this research, especially through the research programs PROFICO and TENAES.
M. De Francia and G. Goya are fellows of CONICET; R. C. Mercader and H.
Vucetich are members of CONICET.

\begin{table}[p]
\caption{\label{TabI} Bounds on the Principle of Equivalence
 	violation parameters.}
\begin{tabular}[p]{|p{2in}ccr|}
\hline
\hline
Experiment	& $\Delta \xi_P$& $\Delta \xi_I$ & Reference\\
\hline
H-maser {\em vs} SCSO	& $2\times10^{-2}$ & --- & \cite{Turneaure}\\
Free neutrons {\em vs}
$^{147}$Sm res\-onan\-ce	& $7\times10^{-3}$ & --- & \cite{OkloRS}\\
``Moving''  {\em vs}
``Rest" $^{57}$Fe	& --- & $5\times10^{-5}$  & \cite{Turner-Hill}\\
Null M\"o\ss{}bauer	&$3.6\times10^{-5}$ & --- &
			   \cite{Red1}\\
Null M\"o\ss{}bauer	&  $3.8\times10^{-5}$  &
			    $5.4\times10^{-8}$ & This paper\\
\hline
\end{tabular}
\end{table}

\end{document}